\documentclass[pdflatex,sn-mathphys-ay]{sn-jnl}

\usepackage{graphicx}%
\usepackage{multirow}%
\usepackage{amsmath,amssymb,amsfonts}%
\usepackage{amsthm}%
\usepackage{mathrsfs}%
\usepackage[title]{appendix}%
\usepackage{xcolor}%
\usepackage{textcomp}%
\usepackage{manyfoot}%
\usepackage{booktabs}%
\usepackage{algorithm}%
\usepackage{algorithmicx}%
\usepackage{algpseudocode}%
\usepackage{listings}%
\usepackage{multirow} 

\theoremstyle{thmstyleone}%
\theoremstyle{thmstyletwo}%

\theoremstyle{thmstylethree}%

\begin{document}

\title[Efficient and scalable clustering of survival curves]{Efficient and scalable clustering of survival curves}

\author*[1,2]{Nora M. Villanueva}\email{nmvillanueva@uvigo.gal}

\author[1,2]{Marta Sesteo}\email{sestelo.gal}

\author[3]{Luís Meira-Machado}\email{lmachado@math.uminho.pt}

%

\affil*[1]{\orgdiv{Department of Statistics and O.R. \& SiDOR Group}, \orgname{Universidade de Vigo}, \orgaddress{\state{36310, Vigo}, \country{Spain}}}

\affil[2]{\orgdiv{Galician Center for Mathematical Research and Technology }, \orgname{CITMAga}, \orgaddress{\state{15782 Santiago de Compostela}, \country{Spain}}}

\affil[3]{\orgdiv{Centre of Mathematics}, \orgname{Universidade de Minho}, \orgaddress{\state{Gumarães}, \country{Portugal}}}


\abstract{Survival analysis encompasses a broad range of methods for analyzing time-to-event data, with one key objective being the comparison of survival curves across groups. Traditional approaches for identifying clusters of survival curves often rely on computationally intensive bootstrap techniques to approximate the null hypothesis distribution. While effective, these methods impose significant computational burdens. In this work, we propose a novel approach that leverages the k-means and log-rank test to efficiently identify and cluster survival curves. Our method eliminates the need for computationally expensive resampling, significantly reducing processing time while maintaining statistical reliability. By systematically evaluating survival curves and determining optimal clusters, the proposed method ensures a practical and scalable alternative for large-scale survival data analysis. Through simulation studies, we demonstrate that our approach achieves results comparable to existing bootstrap-based clustering methods while dramatically improving computational efficiency. These findings suggest that the log-rank-based clustering procedure offers a viable and time-efficient solution for researchers working with multiple survival curves in medical and epidemiological studies.
}




\maketitle

\section{Introduction}
Survival analysis is a fundamental statistical approach used in medical, biological, and epidemiological research to study time-to-event data. A key aspect of survival analysis is the estimation of the survival function, particularly in the presence of right-censored data. The most commonly used non-parametric estimator in such cases is the Kaplan-Meier estimator \citep{kaplan1958nonparametric}, which provides an empirical estimate of the survival probability over time.

In survival studies, researchers are often interested in assessing differences in survival among different groups of participants. For instance, in an observational survival study, comparisons may be made between individuals of different age groups, genders, racial or ethnic backgrounds, or geographical locations. In clinical trials, survival outcomes between participants receiving different treatments, such as a new drug versus a placebo, are frequently analyzed.

Several statistical tests have been developed to compare survival curves, with the log-rank test being the most widely used due to its efficiency in detecting differences in survival distributions \citep{mantel1966evaluation}. The log-rank test evaluates whether survival functions differ significantly between groups under the null hypothesis that they are identical.

When the null hypothesis is rejected, indicating significant differences between survival curves, multiple pairwise comparisons may be required. In R, the \texttt{survminer} package provides the function \texttt{pairwise\_survdiff}, which allows for two-by-two survival curve comparisons. However, as the number of groups increases, multiple testing becomes a challenge, necessitating corrections to control the family-wise error rate.

Commonly used multiple testing correction methods include Bonferroni \citep{dunn1961multiple}, Holm \citep{holm1979simple}, Hochberg \citep{hochberg1988sharper}, and Hommel \citep{hommel1988stagewise}, all of which are implemented in the \texttt{p.adjust function} of the R \texttt{stats} package. For example, when comparing survival curves among 15 groups, such as in a breast cancer dataset analyzed in our application section, the number of pairwise comparisons reaches ${15 \choose 2} = 105$. Conducting multiple hypothesis tests at this scale makes interpretation complex and computationally intensive.

While multiple pairwise comparisons allow for survival curve differentiation, they do not inherently provide a clustering mechanism to group similar curves. The 2019 study by \citet{villanueva2019} introduced a methodology to address this issue by clustering survival curves with an automatic selection of the optimal number of groups, based on resampling techniques. However, the method is computationally demanding due to the extensive use of bootstrap resampling. Although effective, this method introduces substantial computational costs, making it impractical for large-scale applications.

A significant drawback of bootstrap-based clustering methods is the necessity of multiple resampling iterations, where each step involves estimating survival functions, computing test statistics, and evaluating clustering results. This process can be computationally prohibitive, especially when analyzing high-dimensional survival data. To address these limitations, our study proposes a new method that employs the log-rank test as a clustering criterion, thereby significantly reducing computational overhead while maintaining robust statistical properties. Unlike bootstrap-based approaches, the new method provides a direct and computationally efficient means of determining whether two or more survival curves differ significantly. Our method builds upon this principle, systematically forming clusters by iteratively merging survival curves with non-significant log-rank test results, ensuring an optimal partitioning of survival groups.

The remainder of this paper is structured as follows:  Section \ref{methodology} outlines the proposed methodology and its theoretical underpinnings. Section \ref{simulations} presents simulation studies comparing the log-rank-based clustering method with existing bootstrap-based approaches, as well as it is illustrated with the application to a real-world data and evaluated in terms of  computational efficiency. Finally, Section \ref{conclusion} summarizes our findings, discusses potential applications, and outlines directions for future research.

\section{Methodology}
\label{methodology}
This section builds upon the foundational work on survival curve clustering presented in \citet{villanueva2019}. We assume a random censoring model with $J$ samples, where observations from $n_j$ individuals are drawn from population $j$ ($j = 1, \ldots, J$). The total number of observations is denoted as $n = \sum_{j=1}^{J} n_j$, and these observations are assumed to be mutually independent. Let $T_{ij}$ represent the event time for the $i$-th individual in sample $j$, and assume that $T_{ij}$ is subject to a right-censoring variable, $C_{ij}$, which is independent of $T_{ij}$. Due to censoring, instead of observing $T_{ij}$, we observe the pair $(\widetilde{T}_{ij},\Delta_{ij})$, where $\widetilde{T}_{ij}= \min(T_{ij},C_{ij})$ and $\Delta_{ij}=I(T_{ij}\leq C_{ij})$, with $I(\cdot)$ denoting the indicator function.

Since the censoring time is assumed to be independent of the event process, the survival functions $S_j(t) = P(T_j > t)$ can be consistently estimated using the Kaplan-Meier estimator, based on the observed pairs $(\widetilde T_j, \Delta_j)$. Let $t_1<t_2< \ldots <t_{m_j}$, where $m_j \leq n_j$, denote the ordered and distinct failure times of population $j$ ($j=1,\ldots, J$), and let $d_u$ be the number of observed events in population $j$ at time $t_u$. The Kaplan-Meier survival estimator for population $j$ is then given by:
\[
\widehat S_j(t) = \prod_{u:t_u\leq t} \left(1-\frac{d_u}{R_j(t_u)}\right),
\]
where $R_j(t)=\sum_{i=1}^{n_j}I(\widetilde T_{ij} \geq t)$ represents the number of individuals at risk just before time $t$ in population $j$.

If the null hypothesis of equal survival curves, $H_0: S_1 = \ldots = S_J$, is rejected, we introduce a procedure to identify clusters among the survival curves. Specifically, we test whether the populations $\{1, \ldots, J \}$ can be grouped into $K$ clusters $(G_1, \ldots, G_K)$, where $K <J$, such that $S_i = S_j$ for all $i, j \in G_k$, for each $k = 1, \ldots, K$. It is important to note that $(G_1, \ldots, G_K)$ must form a partition of $\{1, \ldots, J\}$, satisfying the conditions $ G_1 \cup \ldots \cup G_K = \{1, \ldots , J \}$ and $ G_i \cap G_j = \emptyset$ for all $ i \neq j \in \{1, \ldots, K \} $.

Given a fixed number of clusters $K$, we propose a procedure to test the null hypothesis $H_0(K)$, which states that there exists at least one partition $(G_1, \ldots, G_K)$ satisfying the above conditions. The alternative hypothesis $H_1(K)$ asserts that, for any partition $G_1, \ldots, G_K$, there exists at least one group $G_k$ in which $S_i \neq S_j$ for some $(i, j) \in G_k$.

To efficiently test $H_0(K)$ while minimizing computational costs, the proposed procedure follows these steps:

\begin{enumerate}
    \item Given the observed data $\{(\widetilde{T}_{ij}, \Delta_{ij})$, $i=1, \ldots, n_j$\}, $j = 1, \ldots, J$, estimate $\hat S_j$ using the Kaplan-Meier estimator on a common time grid.
    
    \item Obtain the optimal partition $(G_1, \ldots, G_K)$ using clustering techniques such as $k$-means or $k$-medians.

    \item For each cluster $k = 1, \ldots, K$, test the equality of survival functions within the group, i.e., $S_i = S_j$ for all $i, j \in G_k$. This step employs a log-rank test or a similar statistical test.\footnote{The current implementation of this method utilizes the \texttt{survminer} package in R, which supports log-rank \cite{mantel1966evaluation}, Gehan-Breslow \cite{gehan1965} \cite{breslow1970}, Tarone-Ware \cite{tarone1977}, Peto-Peto \cite{peto1972}, a modified version of Peto-Peto, and Fleming-Harrington tests.}

    \item Adjust the $K$ p-values obtained in the previous step using multiple testing correction methods.\footnote{Possible corrections include \cite{bonferroni1935statistical}, \cite{holm1979simple}, \cite{hochberg1988sharper}, \cite{hommel1988stagewise}, \cite{benjamini1995controlling} and \cite{benjaminiyukutieli}.}

    \item The final p-value used to test the initial null hypothesis $H_0(K)$ is the minimum of the $K$ adjusted p-values from the previous step.
\end{enumerate}

Similar to bootstrap-based methods, this procedure must be iteratively applied, starting from $K = 1$ and increasing until a certain hypothesis $H_0(K)$ is not rejected. This ensures an automatic determination of the optimal number of survival groups.

\section{Simulation Studies}
\label{simulations}

This section presents application examples to assess and illustrate the practical usefulness of the new method,the new method as well as a  comparison with the method proposed by \citet{villanueva2019} in terms of computational efficiency. First,the new method  is  applied to simulated data across a series  of  experiments, followed by examples using  real datasets, which naturally involve a greater complexity. All computations were performed on a MacBook Air with an M2 processor, 8 cores, 16 GB of RAM, and a solid-state drive.

\subsection{Experiment I}

To evaluate the performance of our method,  we present the results of two simulation scenarios. These scenarios were selected from the work of \citet{villanueva2019} to enable a direct comparison between the bootstrap-based approach and our log-rank-based method in terms of Type I error rate, statistical power, and computational efficiency. 

\begin{figure}[h!]
    \centering
\includegraphics[width=13cm]{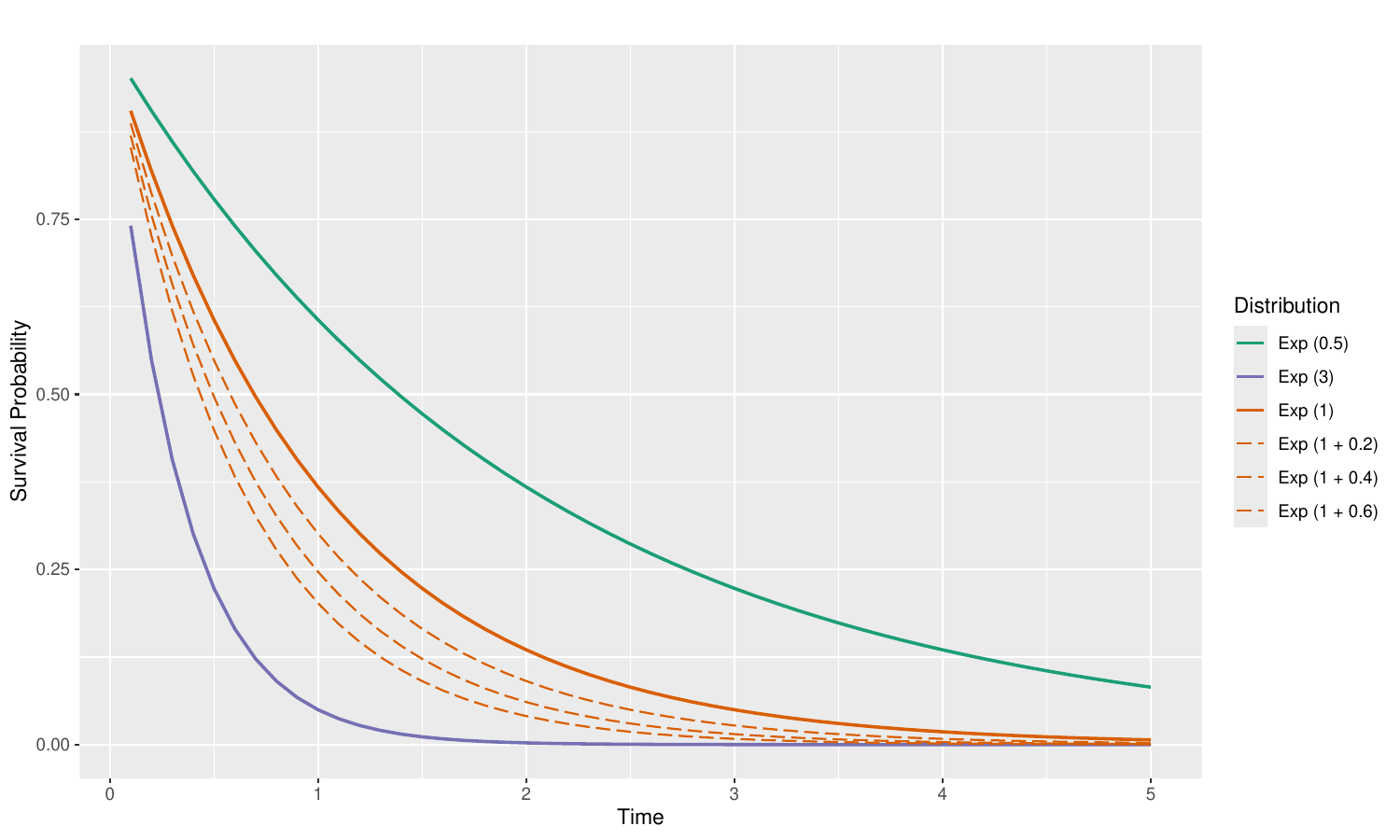}
    \caption{True survival curves of the Experiment Ia. }
    \label{E1}
\end{figure}

The first simulation scenario aims to assess the performance of the proposed procedure by testing a specific null hypothesis $H_0(K)$, where $K = 3$. In this setting, we investigate whether the $J$ survival functions can be classified into three distinct groups. Specifically, we consider a scenario with $J = 6$ populations and the following survival time distributions:  
$F_1, F_2, F_3 \sim Exp(1)$,  
$F_4 \sim Exp(1 + a)$,  
$F_5 \sim Exp(3)$,  
$F_6 \sim Exp(0.5)$,  
where $a$ is a constant. These functions are depicted in Figure~\ref{E1}. The censoring time variable $C$ follows a uniform distribution, $U(0, b)$, with $b$ set to 5 and 3.2, corresponding to censoring rates of 20\% and 30\%, respectively. 

Different values of $a$ were considered, ranging from 0 to 0.6. The case where $a = 0$ corresponds to the null hypothesis, under which all six survival functions can be classified into three groups. When $a \neq 0$, however, the number of distinguishable groups increases to four.

 The Type I error rate and statistical power were estimated based on the empirical rejection proportions obtained from 1,000 repetitions at significance levels of 0.05 and 0.10. We considered different sample sizes $n_j = 50, 100, 200$ for each population $j$. 

Tables~\ref{tab_comp1} and~\ref{tab_comp2} present the empirical Type I error rates and powers for the methods under comparison (for $a = 0.6$ only). Specifically, we compare the following approaches: the  bootstrap-based method ($D_{CM}$) \citep{villanueva2019}, the new method  with Bonferroni correction ($D_{F}$) and the new method without multiple testing correction ($D_{FNC}$).

\begin{table}[ht]
\centering
\caption{Experiment Ia. Estimated Type I error rates for testing $H_0(3)$ using the test statistics $D_{CM}$, $D_{F}$, and $D_{FNC}$, with $J = 6$ populations. Results are presented for two censoring scenarios, where the censoring time variable $C$ follows a uniform distribution: $U(0, 5)$ (approximately 20\% censoring) and $U(0, 3.2)$ (approximately 30\% censoring). \label{tab1}}
\begin{tabular}{rrlllllll}
  \hline
&&&\multicolumn{2}{c}{\textbf{$D_{CM}$}} & \multicolumn{2}{c}{\textbf{$D_{F}$}} & \multicolumn{2}{c}{\textbf{$D_{FNC}$}}   \\ 
  \hline
\textbf{$C$} &  \textbf{$n$} & \textbf{ $\alpha$:} & \textbf{0.05} & \textbf{0.10} & \textbf{0.05} & \textbf{0.10} & \textbf{0.05} & \textbf{0.10} \\ 
  \hline

  & 50 &   &0.06 & 0.09 & 0.05 & 0.08 &0.06 &0.09 \\ 
$U(0,5)$ & 100 &   &0.05 & 0.10 & 0.05 & 0.11 & 0.06&0.11 \\ 
 & 200 &   & 0.05 & 0.09 & 0.06 & 0.10 & 0.06&0.10 \\ 
  \hline
&50 &  &0.05 & 0.08 & 0.05 & 0.07 &0.06 &0.08 \\ 
$U(0,3.2)$ & 100 &  & 0.06 & 0.12 & 0.05 & 0.10 & 0.06&0.11 \\ 
 & 200 &  & 0.05 & 0.10 & 0.05 & 0.10 &0.05 &0.10 \\ 
  \hline
\end{tabular}
\label{tab_comp1}
\end{table}

\begin{table}[ht]
\centering
\caption{Experiment Ia. Rejection probabilities for testing $H_0(3)$ with $a = 0.6$ and $J = 6$ populations, based on the test statistics $D_{CM}$, $D_{F}$, and $D_{FNC}$. Results are reported for two censoring scenarios, where the censoring time variable $C$ follows a uniform distribution: $U(0, 5)$ (approximately 20\% censoring) and $U(0, 3.2)$ (approximately 30\% censoring).}
\label{tab_comp2}
\begin{tabular}{rrlllllll}
  \hline
&&&\multicolumn{2}{c}{\textbf{$D_{CM}$}} & \multicolumn{2}{c}{\textbf{$D_{F}$}} & \multicolumn{2}{c}{\textbf{$D_{FNC}$}}   \\ 
  \hline
\textbf{$C$} &  \textbf{$n$} & \textbf{ $\alpha$:} & \textbf{0.05} & \textbf{0.10} & \textbf{0.05} & \textbf{0.10} & \textbf{0.05} & \textbf{0.10} \\ 
  \hline

 & 50  &  &0.30 & 0.43 & 0.41 & 0.52 &0.47 &0.62 \\ 
$U(0,5)$ & 100   & & 0.74 & 0.85 & 0.87 & 0.92 &0.88 &0.93 \\ 
 & 200   &  &0.99 & 1 & 1 & 1 & 1& 1\\  
  \hline
&50 &    & 0.25 & 0.39 & 0.33 & 0.45 & 0.42& 0.54\\  
$U(0, 3.2)$ & 100   &  & 0.67 & 0.80 & 0.81 & 0.88 &0.83 &0.90 \\  
 & 200   &  &0.97 & 0.99 & 1 & 1 &1 &1 \\  
    \hline
\end{tabular}
\end{table}

When focusing on the first two methods ($D_{CM}$ and $D_F$), both demonstrate good control of the Type I error rate, remaining close to the nominal level across different sample sizes and censoring rates (see Table~\ref{tab_comp1}). As for statistical power, the expected trend is observed: power increases with larger sample sizes and decreases slightly with higher censoring rates (see Table~\ref{tab_comp2}). Notably, the highest power values are attained by the the new method method.

For the third approach (without correction, $D_{FNC}$), the results are very similar to those of the previous methods, suggesting that the correction step has minimal impact in this scenario. This can be explained by the data-generating process: when group assignments are correct, only a single log-rank test is performed within each cluster containing multiple curves, so the multiple testing correction does not substantially affect the outcome. The small differences between $D_{F}$ and $D_{FNC}$ are mainly due to occasional misassignments in smaller sample sizes, where more than one group may contain multiple survival curves.

\begin{figure}[!h]
\centering
\includegraphics[height=17cm]{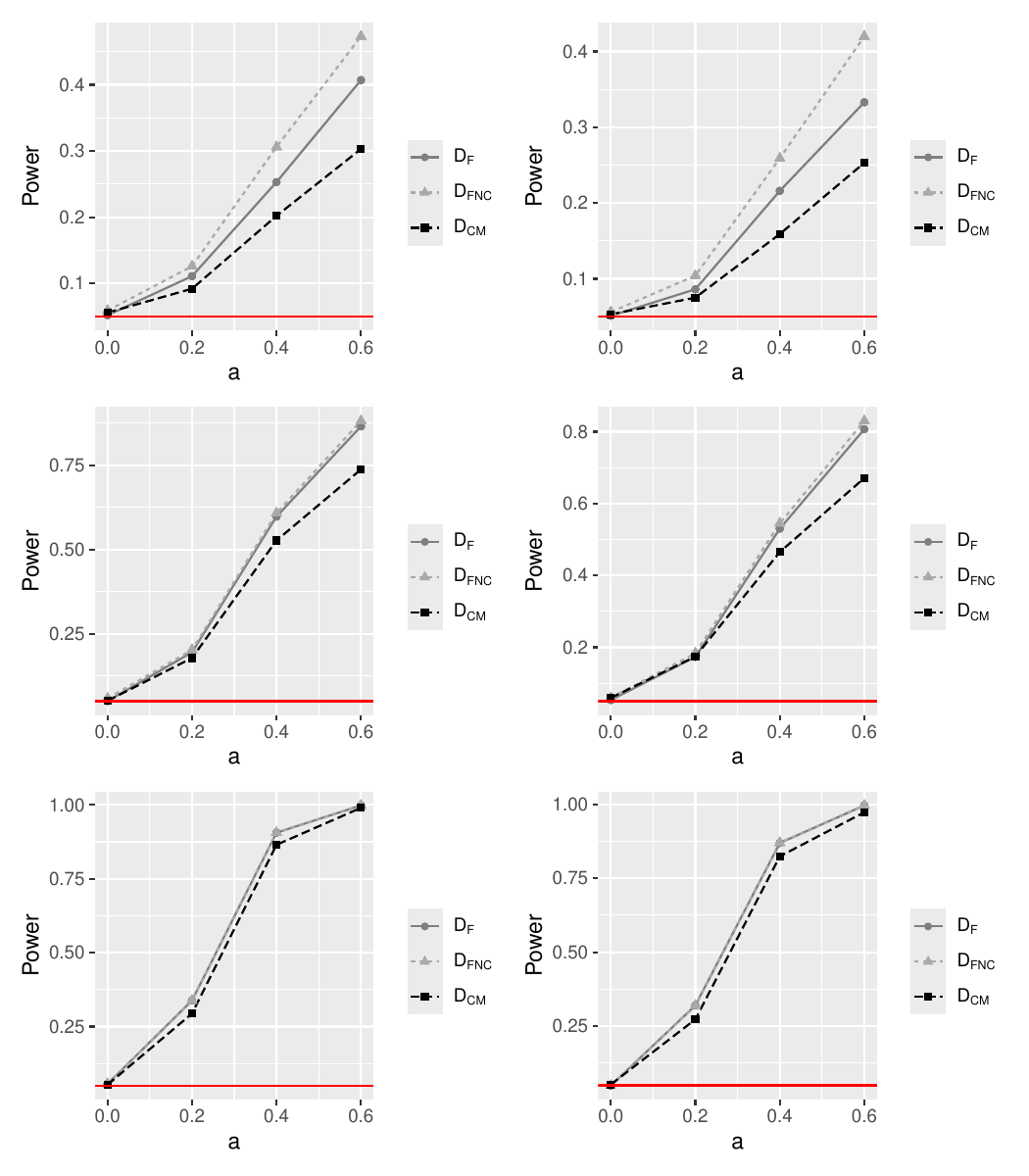}
\caption{Power curves for Experiment Ia. Rejection probabilities of the proposed tests ($D_{CM}$, $D_{F}$, and $D_{FNC}$) at a nominal significance level of 5\% (indicated by the red line). Results are shown for sample sizes $n = 50$, 100, and 200 (top, middle, and bottom panels, respectively), and for censoring time distributions $C \sim U(0, 5)$ (left panels) and $C \sim U(0, 3.2)$ (right panels).}
    \label{power1}
  \end{figure}

Additional power results under alternative hypotheses, as a function of the parameter $a$, are presented in Figure~\ref{power1}. As shown, all three test statistics yield satisfactory power curves, with rejection probabilities increasing as the value of $a$ and the sample size grow. In particular, power approaches 1 for larger sample sizes.

To further investigate the method's behavior when multiple clusters contain more than one survival curve, we modified the previous setting. In this simulation scenario (Ib) , we consider $J = 8$ populations with the following survival time distributions: 
$F_1, F_2, F_3 \sim Exp(1)$,  
$F_4 \sim Exp(1 + a)$,  
$F_5, F_6 \sim Exp(3)$,  
$F_7, F_8 \sim Exp(0.5)$,  
where $a$ is a constant. The censoring variable and other simulation parameters are kept identical to those in the previous scenario. The results are reported in Tables~\ref{tab_comp12} and~\ref{tab_comp22}.

\begin{table}[ht]
\centering
\caption{Experiment Ib. Estimated Type I error rates for testing $H_0(3)$ using the test statistics $D_{CM}$, $D_{F}$, and $D_{FNC}$, with $J = 8$ populations. The censoring time variable $C$ follows a uniform distribution: $U(0, 5)$ (approximately 20\% censoring) and $U(0, 3.2)$ (approximately 30\% censoring).}
\begin{tabular}{rrlllllll}
  \hline
&&&\multicolumn{2}{c}{\textbf{$D_{CM}$}} & \multicolumn{2}{c}{\textbf{$D_{F}$}} & \multicolumn{2}{c}{\textbf{$D_{FNC}$}}   \\ 
  \hline
\textbf{$C$} &  \textbf{$n$} & \textbf{ $\alpha$:} & \textbf{0.05} & \textbf{0.10} & \textbf{0.05} & \textbf{0.10} & \textbf{0.05} & \textbf{0.10} \\ 
  \hline

  & 50 &   &0.06 & 0.10 & 0.06 & 0.10 &0.15 &0.25 \\ 
$U(0,5)$ & 100 &   &0.05 & 0.10 & 0.05 & 0.11 & 0.14&0.28 \\  
 & 200 &   & 0.05 & 0.10 & 0.06 & 0.11 & 0.17&0.28 \\
  \hline
&50 &  &0.06 & 0.09 & 0.05 & 0.09 &0.11 &0.23 \\
$U(0,3.2)$ & 100 &  & 0.04 & 0.11 & 0.05 & 0.09 & 0.15&0.28 \\ 
 & 200 &  & 0.05 & 0.10 & 0.06 & 0.10 &0.15 &0.27\\ 
  \hline
\end{tabular}
\label{tab_comp12}
\end{table}

\begin{table}[ht]
\centering
\caption{Experiment Ib. Rejection probabilities for testing $H_0(3)$ with $a = 0.6$ and $J = 8$ populations, based on the test statistics $D_{CM}$, $D_{F}$, and $D_{FNC}$. The censoring time variable $C$ follows a uniform distribution: $U(0, 5)$ (approximately 20\% censoring) and $U(0, 3.2)$ (approximately 30\% censoring).}
\label{tab_comp22}
\begin{tabular}{rrlllllll}
  \hline
&&&\multicolumn{2}{c}{\textbf{$D_{CM}$}} & \multicolumn{2}{c}{\textbf{$D_{F}$}} & \multicolumn{2}{c}{\textbf{$D_{FNC}$}}   \\ 
  \hline
\textbf{$C$} &  \textbf{$n$} & \textbf{ $\alpha$:} & \textbf{0.05} & \textbf{0.10} & \textbf{0.05} & \textbf{0.10} & \textbf{0.05} & \textbf{0.10} \\ 
  \hline

 & 50  &  &0.26 & 0.39 & 0.38 & 0.47 &0.54 &0.69 \\ 
$U(0,5)$ & 100   & & 0.63 & 0.77 & 0.80& 0.86 &0.90 &0.95 \\   
 & 200   &  &0.97 & 0.99 & 0.99 & 1 & 1& 1\\  
  \hline
&50 &    & 0.22 & 0.36 & 0.32 & 0.41 & 0.46& 0.64\\   
$U(0, 3.2)$ & 100   &  & 0.53 & 0.69 & 0.74 & 0.82 &0.86 &0.92 \\   
 & 200   &  &0.94 & 0.98 & 0.99 & 1 &1 &1 \\   
    \hline
\end{tabular}
\end{table}

When comparing the bootstrap-based method with the new procedure incorporating Bonferroni correction, both approaches approximate the nominal significance level and exhibit increasing rejection probabilities with larger sample sizes. The new method appears to perform best, maintaining Type I error rates closer to the nominal level while achieving higher power for a given sample size.

A critical point to highlight is the impact of omitting multiple testing correction when extracting p-values from the log-rank tests. Unlike the previous experiment, when the curve assignment is correct in this setting, three p-values are produced in step 3 of the procedure, making correction essential. As shown in Table~\ref{tab_comp12}, without correction, the test fails to properly control the significance level.

\subsection{Experiment II}

This simulation experiment aims to evaluate the performance of the complete procedure, in which the null hypothesis $H_0(K)$ is tested sequentially for increasing values of $K$, starting from $K = 1$, until a null hypothesis is not rejected. In this case, we adopt the same scenario described in \citet{villanueva2019}, with $J = 6$ populations and the following survival time distributions: 
$F_1, F_2, F_3 \sim Exp(1)$,  
$F_4, F_5, F_6 \sim Exp(3)$.  
The censoring variable $C$ and other parameters are the same as in the previous experiments.

For the algorithm to function correctly, it must reject the first null hypothesis $H_0(1)$, proceed with further testing, and accept the second hypothesis $H_0(2)$. The results of this simulation are presented in Table \ref{all2}, which reports the proportion of times (in percentage) that the procedure correctly selects the number of clusters $K = 2$ and correctly assigns each population to its respective group. The simulations were conducted using a nominal significance level of 5\%.

As observed, the new method method, when applying multiple testing correction (in this case, Bonferroni correction), performs as expected, achieving a success rate of approximately 95\%, which closely aligns with the theoretical level of $(1 - \alpha)$. This behavior is consistent with the results obtained from the bootstrap-based procedure. Moreover, as in previous experiments, the censoring rate has minimal impact on the methods performance.

\begin{table}[h]
\centering
\caption{Experiment II. Percentage of simulations (out of 1000 repetitions) in which the procedure correctly identifies the number of clusters ($K = 2$) and accurately assigns each population to its corresponding group, using a nominal significance level of 5\%. \label{all2}}
\begin{tabular}{rrllll}
  \hline

\textbf{$C$} &  \textbf{$n$} & & \textbf{$D_{CM}$} & \textbf{$D_{F}$} &\textbf{$D_{FNC}$}  \\ 
  \hline

  & 50 &   &95.3 & 94.4 & 89.8  \\ 
$U(0,5)$ & 100 &   & 96.6& 94.6 & 88.6\\ 
 & 200 &   & 95.2 & 94.4 & 89.9  \\ 
  \hline
&50 &  &95.4 & 94.3 & 89.1  \\ 
$U(0,3.2)$ & 100 &  & 93.2 & 94.4 & 89.7  \\ 
 & 200 &  & 95.3 & 94.4 & 89.3 \\ 
  \hline
\end{tabular}
\end{table}

\subsection{Experiment III}
The final simulation experiment aims to assess the impact of multiple comparison corrections on the performance of the proposed procedure. Addressing multiple comparisons is essential to preserve the validity of statistical inferences. Several correction methods are employed to control error rates, each with distinct properties.

The Bonferroni correction remains one of the most widely used methods for controlling the family-wise error rate (FWER), due to its simplicity: the significance level $\alpha$ is divided by the number of comparisons. While it offers strong protection against false positives, it can be overly conservative, particularly when the number of comparisons is large, often reducing statistical power.

The Holm method, a sequentially rejective extension of Bonferroni, mitigates this conservatism by adjusting rejection thresholds stepwise, thereby improving sensitivity while still maintaining FWER control. The Hommel procedure \citep{hommel1988stagewise}, also considered in our simulations, is another FWER-controlling method that is often less conservative than both Bonferroni and Holm, especially under dependence among tests, thus potentially offering greater power.

Alternatively, the Benjamini-Hochberg (BH) procedure controls the false discovery rate (FDR), making it particularly suitable in contexts where identifying as many true effects as possible is desirable, even at the cost of accepting a controlled proportion of false positives. 

Beyond these, additional correction methods are available through functions such as \texttt{pairwise\_survdiff()} in the \texttt{survminer} package. The optimal choice among these methods depends on the study’s objectives and the acceptable balance between Type I and Type II error rates.

To evaluate the practical impact of these corrections within our framework, we conducted extensive simulation studies under varying sample sizes, effect sizes, and censoring distributions.

In the previous experiments, the survival time distributions were assumed to belong to the same family. To further evaluate the performance of the proposed procedure under more complex and realistic conditions, similar to those encountered in practical data analysis, this final experiment introduces greater variability in distribution types.

Specifically, three different distribution families are considered: Gompertz, Weibull, and Exponential, rather than limiting the analysis to the exponential family alone. In addition, several correction methods are applied to control error rates.

\begin{figure}[h]
    \centering
\includegraphics[width=12cm]{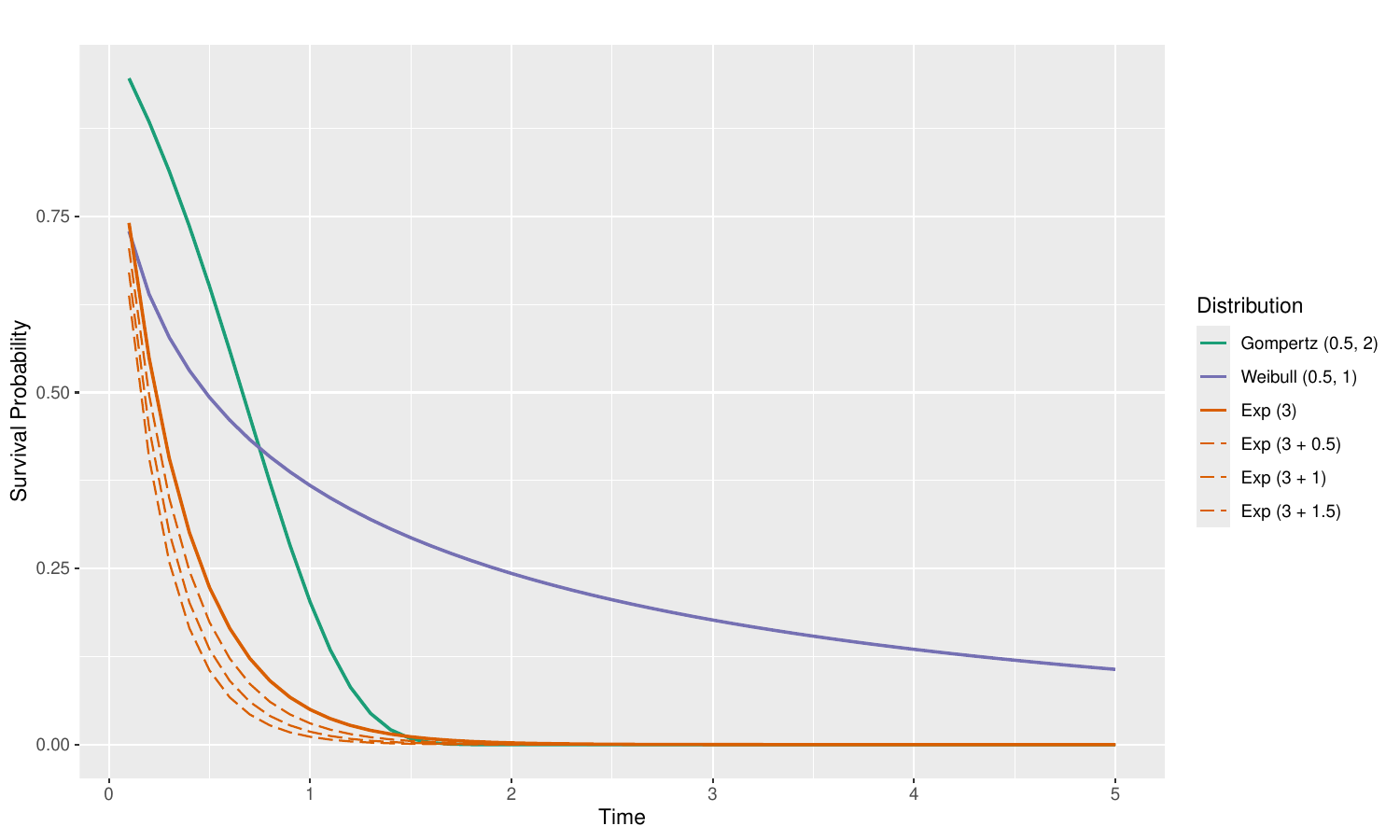}
    \caption{True survival curves of the Experiment III. }
    \label{E3}
\end{figure}

We consider a scenario with $J = 8$ populations and the following survival time distributions: 
$F_1, F_2, F_3 \sim Exp(3)$,  
$F_4 \sim Exp(3 + a)$,  
$F_5, F_6 \sim Gompertz(0.5, 2)$,  
$F_7, F_8 \sim Weibull(0.5, 1)$,
where $a$ is a constant. These functions are shown in Figure~\ref{E3}. The censoring time variable $C$ follows a uniform distribution, $U(0, b)$, with $b$ set to 5 and 3.2, corresponding to 20 percent and 30 percent censoring rates, respectively.

Values of $a$ range from 0 to 1.5. When $a = 0$, the null hypothesis holds, meaning all eight survival functions can be grouped into three clusters. For $a$ different from zero, a fourth group emerges.

The Type I error rate and statistical power were estimated based on empirical rejection proportions obtained from 1,000 repetitions at significance levels of 0.05 and 0.10.
The censoring variable, sample sizes, and all other parameters are the same as those used in the previous experiments.

Tables~\ref{tab_correction} and~\ref{tab_correction_pow} present the results for the simulated Type I error and statistical power, respectively, considering different censoring rates and sample sizes, with a focus on the effect of various correction methods. The results consistently show that the Bonferroni, Holm, BH, and Hommel procedures yield very similar empirical outcomes in terms of both Type I error control and statistical power, across a wide range of sample sizes and censoring conditions.

For example, with $n = 100$ and $a = 1.0$, the four methods produced rejection rates between 0.69 and 0.70 at the 5 percent level, which are practically indistinguishable. This suggests that within the proposed framework, where only a small number of tests is typically performed for each grouping of candidates, the theoretical differences between these correction methods have minimal practical impact. Therefore, any of them can be used without significantly affecting the final clustering outcome.

However, when dealing with a larger number of comparisons, we recommend selecting the correction method based on the study’s tolerance for false positives: the BH procedure is more appropriate when sensitivity is prioritized, while Holm or Bonferroni should be preferred in more conservative scenarios.

Finally, although resampling-based corrections (such as bootstrap or permutation adjustments) could also be considered, their higher computational cost runs counter to the efficiency goals of our method and, in this case, offers no practical benefit.

These findings suggest that, within the structure of our method, where the number of comparisons is relatively small, the choice of correction method has little influence on the final clustering outcome. As a result, any of these approaches can be applied without compromising the validity or performance of the procedure. However, it is important to note that this observation does not extend to settings involving a large number of simultaneous tests, such as omics studies or genome wide survival scans, where differences between family wise error rate and false discovery rate control become more relevant.

In practical terms, we recommend the Benjamini and Hochberg procedure when sensitivity is the primary objective, meaning the goal is to detect as many meaningful differences as possible. If limiting false positives is a greater concern, the Holm method provides a balanced alternative, while the Bonferroni procedure may be preferred in particularly conservative situations. Although corrections based on resampling, such as permutation or bootstrap methods, may offer improved calibration in certain contexts, their computational cost conflicts with the efficiency goals of our clustering algorithm and, based on our results, they are unlikely to yield substantial benefit in this setting.

\begin{table}[ht]
\centering
\caption{Experiment III. Estimated Type I error for testing $H_0(3)$ using the test statistics $D_{FNC}$ (no correction), $D_{F}$ (Bonferroni correction), $D_{H}$ (Holm correction), $D_{BH}$ (Benjamini and Hochberg correction), and $D_{HOM}$ (Hommel correction), with $J = 8$ populations. The censoring time variable $C$ follows a uniform distribution $U(0, 5)$ (approximately 20 percent censoring) and $U(0, 3.2)$ (approximately 30 percent censoring).}
\begin{tabular}{rrlllllllllll}
  \hline
&&& \multicolumn{2}{c}{\textbf{$D_{FNC}$}} &\multicolumn{2}{c}{\textbf{$D_{F}$}} & \multicolumn{2}{c}{\textbf{$D_{H}$}} & \multicolumn{2}{c}{\textbf{$D_{BH}$}} & \multicolumn{2}{c}{\textbf{$D_{HOM}$}}  \\ 
  \hline
\textbf{$C$} &  \textbf{$n$} & \textbf{ $\alpha$:} & \textbf{0.05} & \textbf{0.10} & \textbf{0.05} & \textbf{0.10} & \textbf{0.05} & \textbf{0.10} & \textbf{0.05} & \textbf{0.10} & \textbf{0.05} & \textbf{0.10} \\ 
  \hline

  & 50  &   & 0.148 & 0.260     &0.058 & 0.105 & 0.058 & 0.105 & 0.059 & 0.109 & 0.059 &0.108  \\
$U(0,5)$ & 100 &   & 0.154 &0.284&0.057 & 0.101 & 0.057 & 0.101 & 0.057 & 0.106& 0.057 & 0.105 \\
         & 200 &   &  0.160 & 0.296&0.057 & 0.103 & 0.057 & 0.103 & 0.057 & 0.105& 0.057 & 0.104 \\
  \hline
         & 50  &   & 0.140 & 0.253&0.052 & 0.093 &0.052  & 0.093 & 0.052 & 0.096& 0.052 &0.095 \\
$U(0,3.2)$ & 100 &   & 0.148 & 0.290 &0.062 & 0.108 & 0.062 & 0.108 & 0.064  & 0.108& 0.063 & 0.108 \\
         & 200 &   & 0.160 &0.288&0.052 & 0.103 &  0.052 & 0.103 &  0.053 & 0.106 &  0.053 & 0.105 \\
  \hline
\end{tabular}
\label{tab_correction}
\end{table}

\begin{table}[ht]
\centering
\caption{Experiment III. Rejection probabilities under the alternative hypothesis $H_1$, with $a = 1.5$ and $J = 8$ populations, based on the test statistics $D_{FNC}$ (no correction), $D_{F}$ (Bonferroni correction), $D_{H}$ (Holm correction), $D_{BH}$ (Benjamini and Hochberg correction), and $D_{HOM}$ (Hommel correction). The censoring time variable $C$ follows a uniform distribution $U(0, 5)$ (approximately 20 percent censoring) and $U(0, 3.2)$ (approximately 30 percent censoring).}
\begin{tabular}{rrlllllllllll}
  \hline
&&& \multicolumn{2}{c}{\textbf{$D_{FNC}$}} &\multicolumn{2}{c}{\textbf{$D_{F}$}} & \multicolumn{2}{c}{\textbf{$D_{H}$}} & \multicolumn{2}{c}{\textbf{$D_{BH}$}} & \multicolumn{2}{c}{\textbf{$D_{HOM}$}}  \\ 
  \hline
\textbf{$C$} &  \textbf{$n$} & \textbf{ $\alpha$:} & \textbf{0.05} & \textbf{0.10} & \textbf{0.05} & \textbf{0.10} & \textbf{0.05} & \textbf{0.10} & \textbf{0.05} & \textbf{0.10} & \textbf{0.05} & \textbf{0.10} \\ 
  \hline

  & 50  &   & 0.565 & 0.700     & 0.382 & 0.487 & 0.382& 0.487 &0.387   &0.493 & 0.387 & 0.490  \\
$U(0,5)$ & 100 &   &  0.850 & 0.936 & 0.720 & 0.802 & 0.720& 0.802& 0.722 & 0.809 & 0.722 & 0.807 \\
         & 200 &   & 0.995 & 0.999 & 0.969 & 0.985 & 0.969 &0.985  &  0.972 & 0.987  & 0.972 & 0.987 \\
  \hline
         & 50  &   & 0.542 & 0.687 & 0.353 & 0.479 & 0.353 & 0.479& 0.357 & 0.486 & 0.355 & 0.485 \\
$U(0,3.2)$ & 100 &   & 0.842 & 0.919 & 0.694 & 0.788 & 0.694 & 0.788& 0.701 & 0.792 & 0.701 & 0.791 \\
         & 200 &   & 0.993 & 0.997 & 0.968 & 0.984 & 0.968 & 0.984 & 0.968 & 0.986 &0.968& 0.986 \\
  \hline
\end{tabular}
\label{tab_correction_pow}
\end{table}
\subsection{Real Data Analysis}

Having presented the results for simulated data, we now compare the  new method, based on the log-rank statistic, with  the traditional approach that relies on bootstrap resampling, using a real-world data example that reflect  complex and challenging scenarios. The  data used is the ``rotterdam'' dataset \citep{rotterdam}   which is publicly available in the \texttt{survival} package \citep{survival-package}. t provides essential information on patient outcomes and prognostic factors, making it particularly well suited for survival analysis. It  includes multiple covariates such as tumor characteristics, treatment details, and lymph node involvement, with data collected from a total of 2982 patients. For our analysis, we focus on the time to recurrence (\textit{rtime}) and the number of positive lymph nodes (\textit{nodes}). The variable \textit{rtime} represents the time until recurrence, measured in days, while \textit{recur} is the corresponding censoring indicator, where a value of 1 indicates recurrence and 0 indicates a censored observation.  Among the available covariates, we focus on \textit{nodes}, which represents the number of positive lymph nodes detected. This is a key prognostic factor in breast cancer, as a higher number of affected lymph nodes is strongly associated with disease progression. In our analysis, we categorize \textit{nodes} into 15 levels: 0, 1, 2, up to 13, and an additional category for cases with more than 13 nodes, grouped as level 14.

\begin{figure}[h!]
    \centering
    \includegraphics[width=12cm]{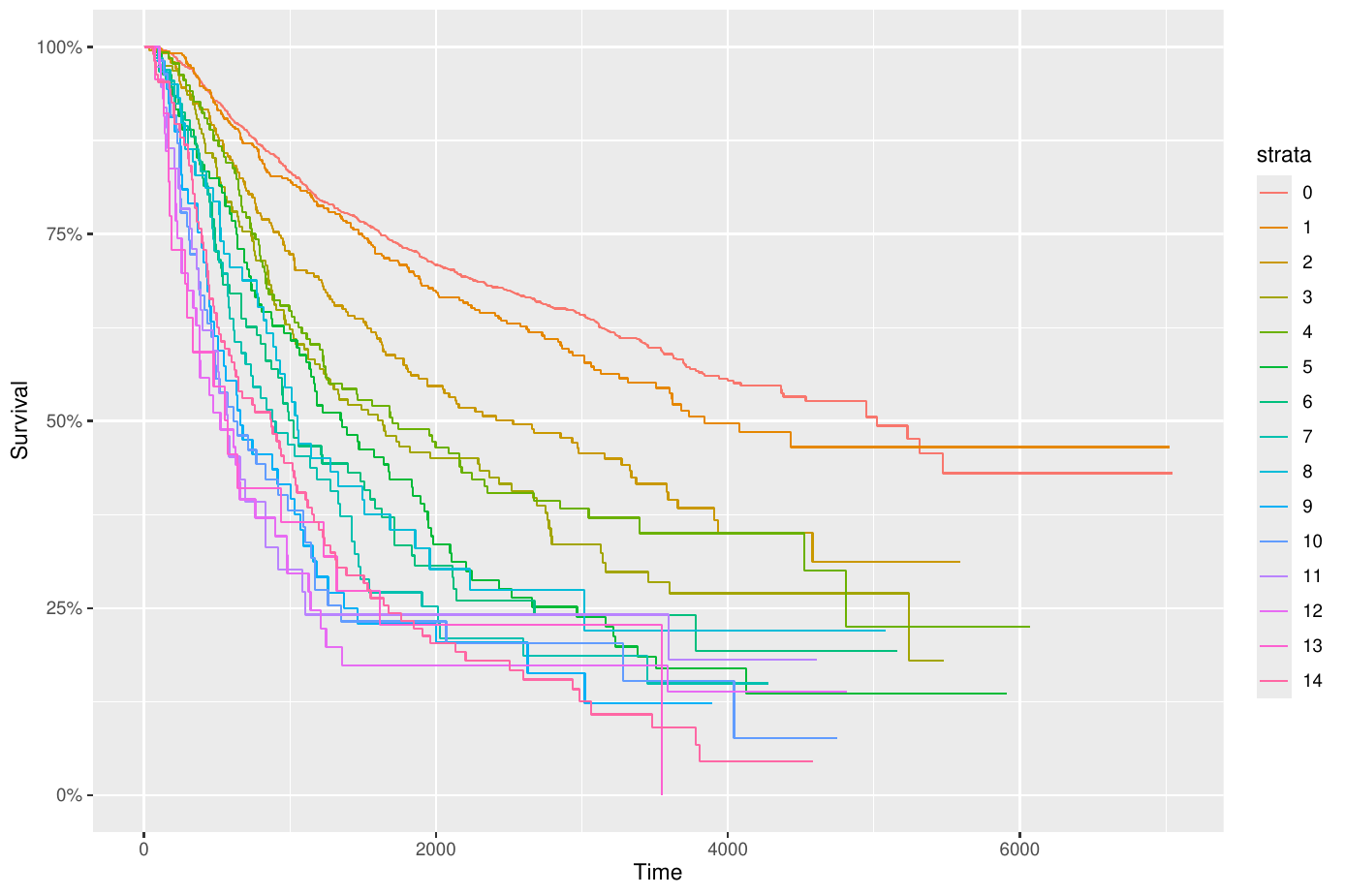}
    \caption{Kaplan-Meier survival curves for recurrence-free survival, stratified by the number of positive lymph nodes.}
    \label{rotter}
\end{figure}

Using the Kaplan Meier estimator, we construct a survival plot with $J = 15$ distinct curves, each representing a different level of lymph node involvement. Figure~\ref{rotter} shows the Kaplan Meier survival curves for recurrence free survival, stratified by the number of positive nodes.

Differences between groups can be assessed using pairwise comparisons, performed with the \texttt{pairwise\_survdiff()} function from the \texttt{survminer} package \citep{Kassambara:2017aa}. However, with $J = 15$ groups, this leads to 105 pairwise comparisons, making direct interpretation difficult. Table~\ref{noramv:pair} presents the p-values from these comparisons, adjusted for multiple testing using the method proposed by Benjamini and Hochberg \citep{benjamini1995controlling}.

Although this approach provides detailed information on group differences, it does not support the identification of clusters of survival curves with similar patterns, which would facilitate a more interpretable analysis. Therefore, alternative clustering methods may be required to extract meaningful insights from the data.

\begin{table}[htbp]
\scriptsize
\resizebox{\columnwidth}{!}{%
\begin{tabular}{c|cccccccccccccc}
     & 0       & 1       & 2       & 3       & 4       & 5       & 6       & 7       & 8       & 9       & 10      & 11      & 12      & 13      \\
\hline
0  & -     & -     & -     & -     & -     & -     & -     & -     & -     & -     & -     & -     & -     & -     \\
1  & 0.162 & -     & -     & -     & -     & -     & -     & -     & -     & -     & -     & -     & -     & -     \\
2  & 0.000 & 0.001 & -     & -     & -     & -     & -     & -     & -     & -     & -     & -     & -     & -     \\
3  & 0.000 & 0.000 & 0.043 & -     & -     & -     & -     & -     & -     & -     & -     & -     & -     & -     \\
4  & 0.000 & 0.000 & 0.195 & 0.617 & -     & -     & -     & -     & -     & -     & -     & -     & -     & -     \\
5  & 0.000 & 0.000 & 0.000 & 0.124 & 0.030 & -     & -     & -     & -     & -     & -     & -     & -     & -     \\
6  & 0.000 & 0.000 & 0.000 & 0.061 & 0.024 & 0.712 & -     & -     & -     & -     & -     & -     & -     & -     \\
7  & 0.000 & 0.000 & 0.000 & 0.008 & 0.002 & 0.186 & 0.431 & -     & -     & -     & -     & -     & -     & -     \\
8  & 0.000 & 0.000 & 0.002 & 0.169 & 0.091 & 0.830 & 0.877 & 0.401 & -     & -     & -     & -     & -     & -     \\
9  & 0.000 & 0.000 & 0.000 & 0.000 & 0.000 & 0.032 & 0.124 & 0.401 & 0.126 & -     & -     & -     & -     & -     \\
10 & 0.000 & 0.000 & 0.000 & 0.000 & 0.000 & 0.028 & 0.088 & 0.362 & 0.105 & 0.982 & -     & -     & -     & -     \\
11 & 0.000 & 0.000 & 0.000 & 0.004 & 0.001 & 0.095 & 0.162 & 0.431 & 0.135 & 0.920 & 0.982 & -     & -     & -     \\
12 & 0.000 & 0.000 & 0.000 & 0.000 & 0.000 & 0.013 & 0.025 & 0.132 & 0.031 & 0.507 & 0.644 & 0.728 & -     & -     \\
13 & 0.000 & 0.000 & 0.000 & 0.004 & 0.002 & 0.061 & 0.141 & 0.412 & 0.156 & 0.830 & 0.830 & 0.830 & 0.903 & -     \\
14 & 0.000 & 0.000 & 0.000 & 0.000 & 0.000 & 0.006 & 0.050 & 0.365 & 0.071 & 0.991 & 0.949 & 0.971 & 0.577 & 0.745 \\
\end{tabular}
} %
\caption{P-values for pairwise comparisons based on the number of positive lymph nodes, using the Benjamini and Hochberg correction for multiple testing.}
\label{noramv:pair}  
\end{table}


Applying the traditional approach based on bootstrap resampling to the ``rotterdam'' dataset, using a significance level of 0.05, $B = 500$ resampling iterations, and the k-means algorithm, the null hypothesis $H_0(1)$ is rejected (p-value $<$ 2e-16), as is $H_0(2)$ (p-value $<$ 2e-16), while $H_0(3)$ is not rejected, with a p-value of 0.056.  This results were obtained by using the  \texttt{clustcurv} package in R \citep{Villanueva2021clustcurvAR}.

\begin{figure}[ht]
    \centering
    \includegraphics[width=12cm]{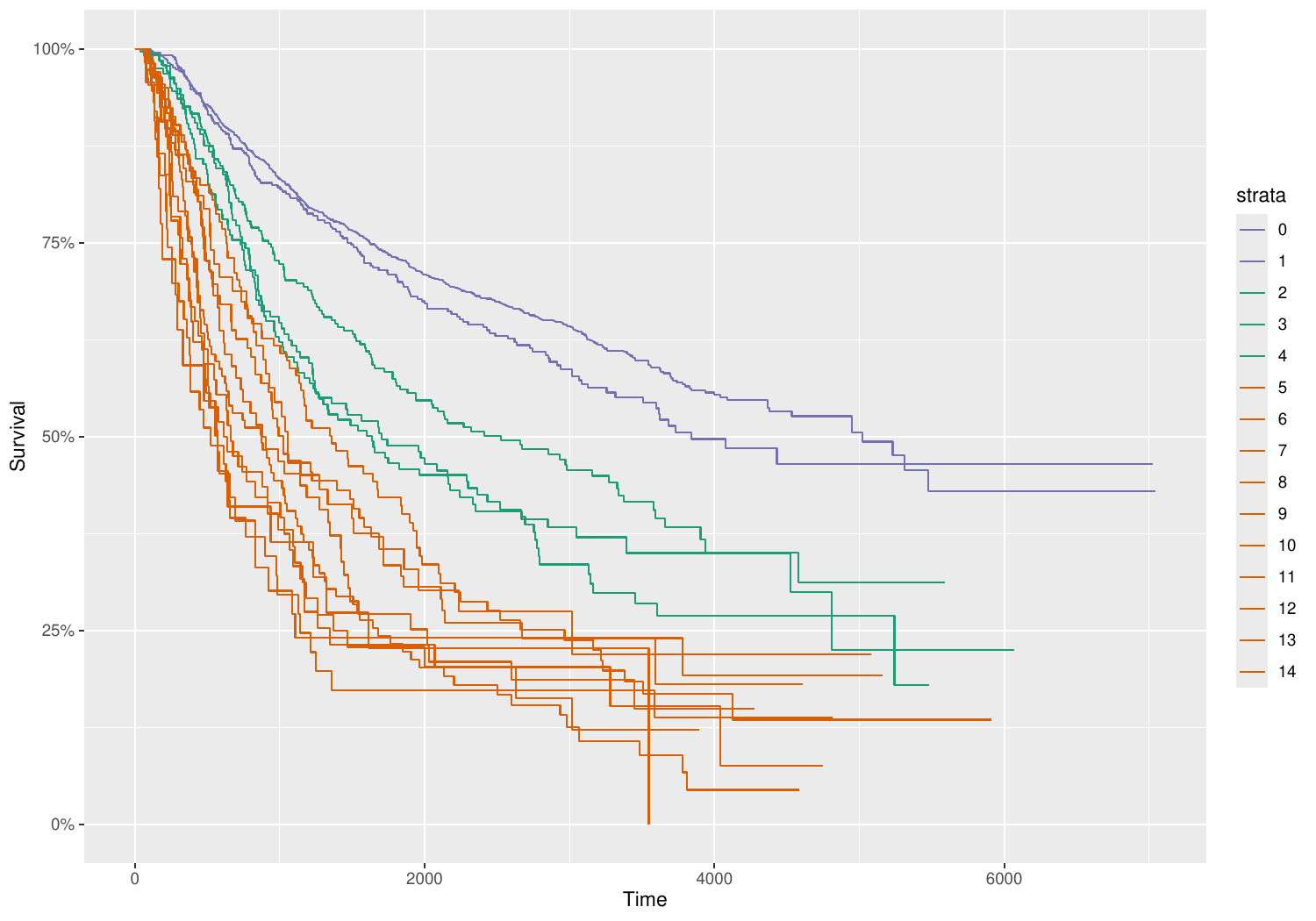}
    \caption{Estimated survival curves stratified by the number of positive lymph nodes. Each curve is colored according to the group to which it belongs (in this case, three groups, $K=3$).}
    \label{rotter_K4}
\end{figure}

The assignment of the survival curves into three distinct groups is shown in Figure~\ref{rotter_K4}. Notably, the presence of one or no positive lymph nodes appears to be associated with a more favorable prognosis. Group 1 includes patients with one or no affected nodes, group 2 includes those with up to four positive nodes, and group 3 includes patients with five or more positive nodes. It is worth noting that the new method  yields the same result regarding the optimal number of groups ($K = 3$) and the assignment  of the curves into their group.

\begin{table}[h!]
\begin{center}
 \caption{Computational execution times in seconds based on a single run.}
\label{noramv:compapp}  
\begin{small}
\begin{tabular}{lllllrr}
\noalign{\smallskip}
\hline\noalign{\smallskip}
dataset  & $J$ & $K$ & $n$ &Method & Parallelization & Time (s)   \\
\hline\noalign{\smallskip}
\multirow{3}{*}{rotterdam} & \multirow{3}{*}{} & \multirow{3}{*}{3} &\multirow{3}{*}{2982} & \citet{villanueva2019}  & FALSE  &  162.20  \\  
          &         15        &                  &  &  \citet{villanueva2019}      &  TRUE  &  58.12  \\ 
          &                    &                  &  &  new method   &  -  &  0.39   \\ 
\hline
\end{tabular}
\end{small}
\end{center}
\end{table}

\section{Discussion}
\label{conclusion}
In this paper, we introduced a fast and scalable method for clustering survival curves that  avoids the use of computationally intensive bootstrap procedures while enabling the automatic identification of survival groups. By leveraging the log-rank test, our method yields results comparable to those reported by \citet{villanueva2019}, but with a significantly reduced computational cost. The proposed method simplifies the detection of distinct survival patterns without compromising reliability or validity, offering a practical and time efficient alternative for analyzing survival data across a range of applications.

Extensive simulation studies show that the method reduces execution time by approximately 98 percent compared to traditional approaches based on bootstrap resampling, making it highly scalable and suitable for large datasets. Importantly, this substantial gain in efficiency does not come at the expense of statistical performance: the Type I error rate remains well controlled, and the method achieves a success rate of around 95 percent, closely matching the expected $(1 - \alpha)$ threshold.

Although the log-rank test is a standard tool for comparing survival curves, its performance is known to be optimal under the assumption of proportional hazards. When this assumption does not hold, alternative approaches may offer better sensitivity to survival differences at specific time intervals. A well known example is the Peto and Peto modification of the Gehan-Wilcoxon test \citep{peto1972}, which places more weight on early events. This makes it particularly suitable for applications where early survival differences are most relevant. Future work could explore how incorporating such alternative weighting schemes might enhance group detection and improve robustness across varied survival scenarios.

Further research will also focus on extending the theoretical properties of the proposed method, particularly in relation to multiple testing corrections, and evaluating its applicability in broader contexts of survival analysis. In addition, a deeper investigation into its behavior under different censoring mechanisms and alternative distance metrics may offer further insights into its practical implementation.

\backmatter

\bmhead{Supplementary information}

For the sake of transparency and reproducibility, the R code for this study can be found in the following GitHub repository: \href{https://github.com/noramvillanueva/fastSCC}{R code Github}. In addition, the  \href{https://CRAN.R-project.org/package=clustcurv}{clustcurv} package is freely available on CRAN, as well as the the two dataset used are publicly available at the  \href{https://CRAN.R-project.org/package=survival}{survival} R package.

\bmhead*{Acknowledgements}

The authors acknowledge financial support by the Spanish Ministry of Science, Innovation and Universities through project PID2023-148811NB-I00 (funded by (AEI/FEDER, UE) and by the Portuguese national funds through FCT --Fundação para a Ciência e a Tecnologia, I.P., under the UID/00013/2025: Centro de Matemática da Universidade do Minho (CMAT/UM) Program Contract, and the project reference 2023.14897.PEX (DOI: 10.54499/2023.14897.PEX).

\bmhead*{Financial disclosure}

None reported.

\bmhead*{Declarations}

The authors declare no potential conflict of interests.

%


\newpage

\begin{appendices}

\section{Bonferroni Correction and Family-Wise Error Rate\label{app1}}

In multiple hypothesis testing, controlling the probability of making at least one Type I error (false positive) is crucial. The Bonferroni correction \citep{bonferroni1935statistical} is a widely used method that ensures control over the Family-Wise Error Rate (FWER). This appendix presents a mathematical derivation of the Bonferroni correction and discusses its implications.

\subsection{Derivation of the Bonferroni Correction\label{app1.1a}}

Let us consider a set of $m$ independent hypothesis tests, each conducted at an individual significance level $\alpha'$. The probability that a single test does not produce a Type I error under the null hypothesis is given by:

\begin{equation*}
P(\text{no Type I error}) = 1 - \alpha'.
\end{equation*}

Assuming independence among the tests, the probability that none of the $m$ tests result in a Type I error is:

\begin{equation*}
P(\text{no Type I errors in } m \text{ tests}) = (1 - \alpha')^m.
\end{equation*}

The probability of making at least one Type I error across the $m$ tests, known as the Family-Wise Error Rate (FWER), is therefore:

\begin{equation*}
FWER = 1 - (1 - \alpha')^m.
\end{equation*}

To ensure that the overall error rate does not exceed a pre-specified level $\alpha$, we impose the condition:

\begin{equation*}
1 - (1 - \alpha')^m \leq \alpha.
\end{equation*}

Solving for $\alpha'$, we obtain:

\begin{equation*}
(1 - \alpha')^m \geq 1 - \alpha.
\end{equation*}

Applying the natural logarithm to both sides:

\begin{equation*}
m \ln(1 - \alpha') \geq \ln(1 - \alpha).
\end{equation*}

From the expansion of \( \ln(1 - x) \) in a Taylor series around \( x = 0 \):

\begin{equation*}
\ln(1 - x) = -x - \frac{x^2}{2} - \frac{x^3}{3} - \frac{x^4}{4} - \dots
\end{equation*}

\noindent it can be seen that for small \( \alpha' \) we can approximate $\ln(1 - \alpha') \approx -\alpha'$, because for small values of \( \alpha' \), higher-order terms \( \mathcal{O}(\alpha'^2) \) become negligible. Thus, keeping only the first-order term, we obtain:

\begin{equation*}
\ln(1 - \alpha') \approx -\alpha'.
\end{equation*}

Substituting this approximation into the inequality gives:

\begin{equation*}
m (-\alpha') \geq \ln(1 - \alpha).
\end{equation*}

This simplification is valid as long as \( \alpha' \) is sufficiently small, ensuring that the discarded higher-order terms have a negligible impact.

\begin{equation*}
-\alpha' m \geq \ln(1 - \alpha).
\end{equation*}

\begin{equation*}
\alpha' \leq \frac{-\ln(1 - \alpha)}{m}.
\end{equation*}

For small values of $\alpha$, using the first-order Taylor expansion approximation $\ln(1 - \alpha) \approx -\alpha$, we obtain:

\begin{equation*}
\alpha' \approx \frac{\alpha}{m}.
\end{equation*}

Thus, the Bonferroni correction prescribes setting the individual test significance level as:

\begin{equation*}
\alpha' = \frac{\alpha}{m}.
\end{equation*}

\subsection{Implications and Limitations}

The Bonferroni correction guarantees strong control of the FWER under arbitrary dependence structures of the test statistics. However, it is known to be conservative, leading to reduced statistical power, especially when $m$ is large. More refined methods, such as the Holm-Bonferroni correction \citep{holm1979simple} or the Benjamini-Hochberg procedure \citep{benjamini1995controlling}, attempt to balance Type I error control with improved sensitivity.

\subsection{Impact on the Power of the Log-Rank Test\label{app1.1b}}

While the Bonferroni correction controls FWER, it does so at the expense of statistical power. The log-rank test statistic follows a chi-square distribution under $H_0$:

\begin{equation*}
Z = \frac{O - E}{\sqrt{V}} \sim \mathcal{N}(0,1),
\end{equation*}

where $O$ and $E$ denote observed and expected event counts, respectively, and $V$ represents the variance. Without correction, the rejection threshold corresponds to the quantile:

\begin{equation*}
Z_{\alpha} = \Phi^{-1}(1 - \alpha),
\end{equation*}

where $\Phi^{-1}$ is the inverse standard normal cumulative distribution function. After Bonferroni correction, the adjusted threshold becomes:

\begin{equation*}
Z_{\alpha'} = \Phi^{-1}(1 - \alpha').
\end{equation*}

Since $\alpha' = \frac{\alpha}{m}$, it follows that:

\begin{equation*}
Z_{\alpha'} > Z_{\alpha}.
\end{equation*}

This stricter rejection threshold results in an increased probability of failing to reject false null hypotheses, thereby reducing the test’s statistical power. In particular, for a given effect size $\delta$, the power function of the log-rank test is given by:

\begin{equation*}
1 - \beta = P\left(Z > Z_{\alpha'} \mid H_A\right),
\end{equation*}

where $\beta$ denotes the Type II error rate. Since $Z_{\alpha'}$ increases with $m$, the power of the test decreases as the number of comparisons grows.

\subsection{Asymptotic Behavior and Alternative Approaches\label{app1.1c}}

In large-scale survival analyses where $J$ is large, the Bonferroni correction may become overly conservative, leading to a significant reduction in power. As $J \to \infty$, the number of pairwise comparisons grows quadratically, causing $\alpha'$ to shrink towards zero. Consequently, unless sample sizes grow proportionally to $J$, the probability of detecting true survival differences vanishes asymptotically.

Alternative multiple testing corrections, such as Holm-Bonferroni \citep{holm1979simple} or Benjamini-Hochberg \citep{benjamini1995controlling}, offer more balanced control between FWER and statistical power. While Holm’s method maintains strict FWER control with improved power over Bonferroni, the Benjamini-Hochberg procedure instead controls the False Discovery Rate (FDR), allowing for greater sensitivity in large-scale survival studies.

\end{appendices}

\bibliography{ref}  


\end{document}